\documentclass[11pt,a4paper]{article}
\pdfoutput=1
\usepackage{jcappub}
\usepackage[english]{babel}
\usepackage{graphicx}
\usepackage[margin=1.0in]{geometry}
\usepackage{booktabs}
\usepackage{array}
\usepackage{etoolbox}
\usepackage{multirow}

\newcommand{\planck}{\textit{Planck}}
\newcommand{\neff}{N_\mathrm{eff}}
\newcommand{\LCDM}{$\Lambda \mathrm{CDM}$}
\newcommand{\GeV}{\mathrm{GeV}}
\newcommand{\eV}{\mathrm{eV}}

\newcommand{\Mpc}{\mathrm{Mpc}}
\newcommand{\SLCDM}{S$\Lambda \mathrm{CDM}$}
\newcolumntype{S}{>{\centering\arraybackslash}m{\dimexpr.18\linewidth-8\tabcolsep}}
\newcolumntype{N}{>{\centering\arraybackslash}m{\dimexpr.23\linewidth-8\tabcolsep}}
\newcolumntype{M}{>{\centering\arraybackslash}m{\dimexpr.25\linewidth-7\tabcolsep}}
\newcolumntype{L}{>{\centering\arraybackslash}m{\dimexpr.27\linewidth-7\tabcolsep}}
\newcolumntype{C}{>{\centering\arraybackslash}m{\dimexpr.27\linewidth-1.2\tabcolsep}}
\newcolumntype{G}{>{\centering\arraybackslash}m{\dimexpr.5\linewidth-0.1\tabcolsep}}
\AtBeginEnvironment{bmatrix}{\setlength{\arraycolsep}{10pt}}
\begin{document}
\title{Cosmic microwave background constraints on secret interactions among sterile neutrinos}

\author[a,b]{Francesco Forastieri,}
\author[b]{Massimiliano Lattanzi,}
\author[c]{Gianpiero Mangano,}
\author[d,e]{Alessandro Mirizzi,}
\author[a,b]{Paolo Natoli,}
\author[f]{Ninetta Saviano}
\affiliation[a]{Dipartimento di Fisica e Scienze della Terra, Universit\`a di Ferrara, Via Giuseppe Saragat 1, I-44122 Ferrara, Italy.}
\affiliation[b]{Istituto Nazionale di Fisica Nucleare, Sezione di Ferrara, Via Giuseppe Saragat 1, I-44122 Ferrara, Italy.}
\affiliation[c]{Istituto Nazionale di Fisica Nucleare, Sezione di Napoli, Complesso Univ. Monte S.Angelo, I-80126 Napoli, Italy.}
\affiliation[d]{Dipartimento Interateneo di Fisica ``Michelangelo Merlin,'' Via Amendola 173, 70126 Bari, Italy.}
\affiliation[e]{Istituto Nazionale di Fisica Nucleare - Sezione di Bari, Via Amendola 173, 70126 Bari, Italy.}
\affiliation[f]{ PRISMA Cluster of Excellence and Mainz Institute for Theoretical Physics, Johannes
Gutenberg-Universit\"at Mainz, 55099 Mainz, Germany}
\emailAdd{francesco.forastieri@unife.it}
\emailAdd{lattanzi@fe.infn.it}
\emailAdd{mangano@na.infn.it}
\emailAdd{alessandro.mirizzi@ba.infn.it}
\emailAdd{natoli@fe.infn.it}
\emailAdd{nsaviano@uni-mainz.de}

\abstract{ 
Secret contact interactions among eV sterile neutrinos, mediated by a massive gauge boson $X$ (with $M_X \ll M_W$), and characterized by a gauge coupling $g_X$, have been proposed as a mean to reconcile cosmological observations and short-baseline laboratory anomalies. We constrain this scenario using the latest Planck data  on Cosmic Microwave Background anisotropies, and measurements of baryon acoustic oscillations (BAO). 
We consistently include
the effect of secret interactions on cosmological perturbations, namely the increased density and pressure fluctuations in the neutrino fluid, 
and still find a severe tension between the secret interaction framework and cosmology.
In fact, taking into account neutrino scattering via secret interactions, we derive our own mass bound on sterile neutrinos and find (at 95 \% CL) $m_s < 0.82$~eV or $m_s < 0.29\,\eV$ from Planck alone or in combination with BAO, respectively. These limits confirm  the discrepancy
with the laboratory anomalies. Moreover, we constrain, in the limit of contact interaction, the effective strength $G_X$ to be $ < 2.8 (2.0) \times 10^{10}\,G_F$ from Planck (Planck+BAO). This result, together with the mass bound, strongly disfavours the region with $M_X \sim 0.1$~MeV and relatively large coupling $g_X\sim 10^{-1}$, previously indicated as a possible solution to the small scale dark matter problem. 
}
\maketitle

\section{Introduction}

In recent years there has been a renewed interest towards light sterile neutrinos, suggested by different anomalies observed 
in short-baseline (SBL) neutrino experiments (see \cite{Abazajian:2012ys,Kopp:2013vaa,Gariazzo:2015rra,Gariazzo:2017fdh}  for  recent reviews). 
In particular, these anomalies can be explained postulating a sterile neutrino with mass $m_s \simeq {\mathcal O}$ (1~eV) and active-sterile mixing angle $\theta_s \simeq 0.1$. 
For these values of the sterile neutrino parameters, the new states would be copiously produced in the Early Universe, resulting in a conflict with existing cosmological bounds on primordial radiation density and neutrino mass \cite{Hamann:2011ge,Hannestad:2012ky,Mirizzi:2013gnd}.
For this scenario to survive, a mechanism must be in place to suppress sterile neutrino abundance in the early universe: e.g., large primordial neutrino asymmetries \cite{Hannestad:2012ky,Mirizzi:2012we,Saviano:2013ktj}, free primordial power-spectrum \cite{Gariazzo:2014dla} or low reheating temperature \cite{Yaguna:2007wi}.
Recently, a new mechanism has been proposed, which achieves such a suppression postulating secret interactions among sterile neutrinos, mediated by a massive gauge boson $X$, with $M_X \ll M_W$ \cite{Hannestad:2013ana,Dasgupta:2013zpn,Bringmann:2013vra} 
(see \cite{Archidiacono:2015oma,Archidiacono:2016kkh} for the case of sterile neutrinos interacting with a light pseudoscalar). These secret interactions are described by the following Lagrangian
\begin{equation}
{\mathcal L}= g_X {\bar \nu_s}\gamma_\mu \frac{1}{2}(1-\gamma_5) \nu_s X^{\mu}  \,\ ,
\end{equation}
where $g_X$ is the gauge coupling.
Secret interactions would generate a large matter term in the sterile neutrino sector, that  reduces the effective mixing angle, suppressing the active-sterile oscillations.
Since the secret interactions are confined to the sterile sector, at the beginning this scenario seemed unconstrained. However, it was later realized that as the matter potential generated by the secret coupling declines as the Universe expands, sterile neutrinos would eventually encounter a resonance, when the matter potential becomes of the order of the neutrino vacuum oscillation frequency. This would allow for a sterile neutrino production through the combination of resonant Mikheyev-Smirnov-Wolfenstein  effect \cite{Mikheev:1986wj,Wolfenstein:1977ue} and of non-resonant production  via the secret collisions \cite{Stodolsky:1986dx}. In this regard, it has been shown by some of us that for a coupling constant $g_X \gtrsim 10^{-2}$ and masses of the mediator $M_X \gtrsim {\mathcal O}$(10~MeV) the sterile neutrino production would occur before neutrino decoupling ($T \gtrsim$ 0.1--1~MeV) impacting the yield of light elements during Big Bang Nucleosynthesis (BBN) \cite{Saviano:2014esa}.  For smaller values of the mediator mass, BBN would be unaffected. However, in this case, sterile neutrinos would still be produced at $T \ll$ 0.1~MeV, when the matter potential becomes smaller than the vacuum oscillation term. Even assuming a negligible resonant production, sterile neutrinos would be copiously produced by the collisional term in the secret sector seeded by vacuum mixing, analogously to the 
Dodelson-Widrow mechanism acting for dark matter sterile neutrinos \cite{Dodelson:1993je}. 
In \cite{Mirizzi:2014ama} some of us have shown that this \emph{decoherent} sterile neutrino production would quickly lead to equilibrium among active and sterile species, leading to a sizeable abundance of the latter in conflict with the cosmological neutrino mass bound. In addition, this mechanism reduces the effective number of neutrinos to $N_{\rm eff} \simeq 2.7$ at matter-radiation equality.
For $M_X \gtrsim 0.1$ MeV, sterile neutrinos would be free-streaming before becoming non-relativistic and they would affect the structure formation at scales smaller
than the free-streaming length. Conversely, for masses $M_X \lesssim 0.1$ MeV, as noticed in \cite{Mirizzi:2014ama,Chu:2015ipa}, sterile neutrinos would be at the border between free-streaming and collisional regime at the photon decoupling, so one cannot naively apply the mass constraints as we did before. This range of the parameter space for the secret interactions is particularly interesting since it was previously shown \cite{Dasgupta:2013zpn,Bringmann:2013vra} to have potentially important consequences for the small scale structure of dark matter if the mediator $X$ couples also to dark matter.
Furthermore, possible signatures of secret interactions in the observations
of very-high-energy neutrinos by Icecube has been analyzed in \cite{Cherry:2014xra,Cherry:2016jol}
(see also \cite{Ng:2014pca,Ioka:2014kca,Blum:2014ewa} for recent studies on the impact of secret interactions among active neutrinos on Icecube observations).

In the present work, we pursue a dedicated investigation of $M_X \lesssim 0.1$ MeV  region obtaining constraints by the latest Planck data on the cosmic microwave background. 
The plan of this paper is as follows. In Sec.~2 we discuss the production mechanism of sterile neutrinos associated with secret interactions in the post-decoupling epoch
and we present the existing cosmological bounds on this scenario. In Sec.~3 we present the results of our analysis and we draw our conclusions in Sec.~4.

\section{Secret interaction framework}

\subsection{Sterile neutrino production}

The 3+1 active-sterile neutrino mixing scenario involves  3 active families and 
a sterile species.  Describing  the neutrino system in terms of  $4\times 4$ density matrices $\rho= \rho(p)$, 
the active-sterile flavour evolution  is ruled by the kinetic equations \cite{Dolgov:2002ab}
\begin{equation}
{\rm i}\,\frac{d\rho}{dt} =[{\sf\Omega},\rho]+ C[\rho]\, ,
\label{drhodt}
\end{equation}
see~\cite{Mirizzi:2012we} for a detailed treatment. 
The first term on the right-hand side of Eq.\ (\ref{drhodt}) describes the flavour oscillations Hamiltonian, given by
\begin{eqnarray}
{\sf\Omega}&=&\frac{{\sf M}^2}{2 p} +
\sqrt{2}\,G_{\rm F}\left[-\frac{8  p}{3 }\, \bigg(\frac{{\sf E_\ell}}{M_{\rm W}^2} + \frac{{\sf E_\nu}}{M_{\rm Z}^2}\bigg)
\right] \nonumber \\
&+& \sqrt{2}\,G_{\rm X}\left[-\frac{8  p  {\sf E_s}}{3 M_{\rm X}^2}
\right]  \,\ ,
\label{eq:omega}
\end{eqnarray}
where ${\sf M}^2$ = ${\mathcal U}^{\dagger} {\mathcal M}^2 {\mathcal U}$ is the neutrino mass matrix in flavour basis, 
with ${\mathcal U}$  the active-sterile vacuum mixing matrix.  The terms proportional to the Fermi constant $G_F$ in Eq.~(\ref{eq:omega})  are the standard matter effects in active neutrino oscillations,
while the term proportional to $G_X$   represents the new matter secret potential. 
In particular,  ${\sf E_\ell}$ is related to the energy density of $e^{-}$ and $e^+$ , 
${\sf E_\nu}$  is  the $\nu$-$\nu$ interaction term  proportional to a primordial neutrino asymmetry (that here we assume negligible),
while ${\sf E_s}$ is the energy density associated with $\nu_s$.  
The last term in the right-hand side of Eq.~(\ref{drhodt}) is the collisional integral given by the sum of the standard  ($\propto G_F^2$) and the secret one ($\propto G_X^2$).
Since the flavour evolution typically occurs at neutrino temperature
$T_\nu \ll M_X$ we can   reduce the secret interaction to a  contact form, with an effective
strength 
\begin{equation}
G_X = \frac{\sqrt{2}}{8} \frac{g_X^2}{M_X^2} \,\ . 
\end{equation}

The strong collisional effects produce a damping of the resonant transitions and would bring the system
towards the flavour equilibrium among the different neutrino species  with
a production rate given by  \cite{Kainulainen:1990ds,Mirizzi:2014ama}
\begin{equation}
\Gamma_t \simeq \langle P(\nu_\alpha \to \nu_s) \rangle_{\rm coll} \Gamma_X \,\, ,
\label{eq:gammat}
\end{equation}
where $\langle P(\nu_\alpha \to \nu_s) \rangle_{\rm coll}$ is the average probability of conversions among an active $\nu_{\alpha}$
and a sterile  neutrino $\nu_s$ in a scattering time scale $(\Gamma_X)^{-1}$, where the scattering rate is given by 
\begin{equation}
\Gamma_X \simeq G^2_X T^5_\nu \frac{p}{\langle p \rangle} \frac{n_s}{n_a} \,\  .
\label{eq:gammax}
\end{equation}
In Eq.~(\ref{eq:gammax})  $\langle p \rangle \simeq 3.15 T_\nu$ is the 
average-momentum for a thermal Fermi-Dirac distribution, and $n_s$ and $n_a$ the sterile and active neutrino abundance, respectively.

\subsection{Cosmological bounds: state-of-the-art}
Since the search for sterile neutrinos
in laboratory experiments is presently open, it is important to use as
many observations as possible to corner sterile neutrinos and in particular their production through secret interactions.
 In this context, cosmological observations represent a valid complementary tool
to probe this scenario, being sensitive to the number of neutrinos, to
their mass and to their free streaming characteristic.

In this Section we present the cosmological bounds obtained so far. 
For a coupling constant $g_X \gtrsim 10^{-2}$ and masses of the mediator $M_X \gtrsim {\mathcal O}$(10~MeV)
 the sterile neutrino production would occur before neutrino decoupling ($T \gtrsim$ 0.1--1~MeV).
At this regard,  in \cite{Saviano:2014esa}   it  has been computed the sterile neutrino production 
relevant for BBN. The standard BBN dynamics is altered both by a larger value of $N_{\rm eff}$ and by the spectral distortion of $\nu_e$ when oscillations occur close to the neutrino decoupling.
 Using the present determination of  deuterium primordial abundances,  it was found  that  the $^2$H/H density ratio excludes much of the parameter space at 3$\sigma$, in particular  masses $M_X \geq 40~\textrm{MeV}$ are excluded.
 
For smaller values of the mediator mass  a large matter potential is generated suppressing the sterile neutrino production before the neutrino decoupling. With this choice of parameter ranges, BBN is left unchanged and gives no bound on the model. However,  at lower temperatures when active-sterile oscillations are no longer matter suppressed, sterile neutrinos are still in a collisional regime, due to their secret self-interactions. The interplay between vacuum oscillations and collisions leads to a scattering-induced decoherent production of sterile neutrinos with a fast rate given in Eq.~(\ref{eq:gammat}). 
 At this regard, in \cite{Mirizzi:2014ama} were neglected   the resonant matter effects in the sterile neutrino production, reducing the average probability   in Eq.~(\ref{eq:gammat}) to a pure vacuum one, i.e.
\begin{equation}
\langle P(\nu_\alpha \to \nu_s) \rangle_{\rm coll} \simeq \frac{1}{2} \sin^2 \theta_{\alpha s} \,\ .
\end{equation}
Taking as representative mixing angle $\sin^2 2\theta_{es} \simeq 0.12$ \cite{Giunti:2013aea}, one would expect a sterile neutrino abundance, $n_s \simeq 0.06 \,\ n_a$.
This seemingly negligible population is enough to generate a large scattering rate in the post-decoupling epoch for a sufficiently large $G_X$.
In particular for $G_X \gtrsim 10^8 G_F$ the scattering rate at $T_\gamma \sim 10^{-2}$~MeV would be much larger than the Hubble rate $H$. 
This would lead a fast flavour equilibration between the three active and the sterile species, leading from an initial abundance
\begin{equation}
(n_e, n_\mu, n_\tau, n_s)_{\rm initial} = (1,1,1,0) \,\ ,
\end{equation}
to a final one:
\begin{equation}
(n_e, n_\mu, n_\tau, n_s)_{\rm final} = \left(\frac{3}{4}, \frac{3}{4}, \frac{3}{4}, \frac{3}{4} \right) \,\ ,
\end{equation}
for all the parameters associated with eV sterile neutrino anomalies.

Soon after $\nu_s$ are produced via oscillation, active and sterile neutrinos have a shared grey-body distribution, namely a Fermi-Dirac function 
weighted by a factor 3/4 for each species. However, in the presence of strong secret interactions, these grey-body distributions will fastly evolve
towards a  Fermi-Dirac equilibrium function. The constant number density
(or entropy) constraint implies that the temperature of this final spectrum is reduced by a factor $(3/4)^{1/3}$ with respect to
the initial active neutrino temperature $T_\nu= (4/11)^{1/3} T_\gamma$.  As a consequence of this effect,
the total energy density stored in active and sterile neutrinos is reduced and the value of the effective number of neutrino species decreases down to
$N_{\rm eff} \sim 2.7$ for relativistic neutrinos. 
A further slight reduction would occur 
 at the matter radiation equality, i.e. for $T_{\gamma} \sim 0.7$~eV since eV sterile neutrinos would not be fully relativistic.  
 
Secret interactions also affect the evolution of perturbations in the sterile neutrino fluid. In fact, if sterile states scatters via secret interactions, the free streaming regime is delayed until the scattering rate becomes smaller than the Hubble parameter. 
It means that if $G_X$ is large enough so that this condition holds at the non relativistic transition, sterile neutrinos would never have a free streaming phase, 
but always diffuse \cite{Mirizzi:2014ama,Chu:2015ipa}.
One can obtain the smaller value of $G_X$ for which this occurs comparing the scattering rate with the Hubble rate $H$ at a temperature
$3.15 T_\nu \sim \langle p \rangle \sim m_{\rm s}$
\begin{equation}
G^2_X T^5_\nu \sim H(T_\gamma) \,\ .
\end{equation}
Writing the Hubble rate for $T_\gamma \sim $eV and using that  $T_\nu= (4/11)^{1/3}  (3/4)^{1/3} T_\gamma$
one obtains:
\begin{equation}
G_X \sim 10^{10} G_F \,\ ,
\end{equation}
which corresponds to $M_X \sim 10^{-1}$ MeV for $g_X \simeq 10^{-1}$.

In Ref. \cite{Mirizzi:2014ama}, sterile neutrino production through secret collisions has been examined also in view of the cosmological bounds
on the sterile neutrino mass. Since for $G_X \lesssim 10^{10} G_F$ massive sterile neutrinos are free-streaming at the transition
to the nonrelativistic regime, the peculiar effect of suppression of small-scale matter perturbations, induced by the presence 
of a light, free-streaming species, is preserved. In this case it is legitimate to expect that the mass bounds for non-interacting neutrinos
coming from Cosmic Microwave Background (CMB) and Large Scale Structure (LSS) observations still apply also in the interacting case, as they basically rely on the effect of sterile neutrinos
on the perturbation evolution. With this assumption, it was argued in Ref. \cite{Mirizzi:2014ama} that an eV-mass sterile state,
as suggested by the short baseline laboratory anomalies, was  in tension, at least at the $2\sigma$ level with cosmological bounds available at the time, including those from the 2013 data release of the \planck\ satellite \cite{Ade:2013zuv}. On the contrary, if the coupling is extremely large ($G_X > 10^{10} G_F$), the
free-streaming regime is reached only after the nonrelativistic transition, and the cosmological mass bounds possibly do not apply. 

In this work, we aim to perform a dedicated, self-consistent analysis to derive observational bounds
on secret contact interactions using the latest public Planck data on CMB anisotropies, as well as information from baryon acoustic oscillations (BAO).
In order to do that, we do not rely on mass bounds obtained with the standard assumption of free-streaming neutrinos, but instead
derive our own bounds, taking into account neutrino scattering via secret interactions. Moreover, we also take into account another effect induced by secret interactions, namely the increased density and pressure perturbations in the neutrino fluid,
due to the fact that collisions cause power to flow towards the lower moments of the Boltzmann hierarchy. This produces 
a distinctive signature in the CMB anisotropy spectrum, as we shall see in the next section. Both these improvements
are obtained by considering a collision term, proportional to the scattering rate, in the Boltzmann hierarchy for neutrinos,
and by performing a fully consistent analysis in which both the sterile neutrino mass and the effective strength
are treated as free parameters.

\subsection{Secret interactions and cosmological perturbations}

In this Section we describe the effect of interactions among neutrino species on the evolution of cosmological perturbations, that in turn
determine the observational signatures on the CMB anisotropies and on large scale structures. Writing the sterile neutrino distribution function $f$
as the sum of a zero-th order part $f_0$ and a perturbation $\delta f \equiv f_0 \Psi$, the latter evolves according to the Boltzmann equation:
\begin{equation}
\hat L[\delta f] = \hat C[\delta f]\, ,
\label{eq:boltZ_eq}
\end{equation}
where $\hat L$ is the Liouville operator. The collision term $\hat C$ in the right-hand side takes into account the effect of secret interactions. In principle, the collision term
is a complicated integral involving the matrix elements for the relevant processes;
computing exactly the collision integral is a numerically demanding task, beyond the scope of our work (see e.g. Refs.~\cite{Oldengott:2014qra,Oldengott:2017fhy} for a detailed study of this topic).
Fortunately it is enough, for the purpose of studying the evolution of cosmological perturbations, to resort to the so-called relaxation time approximation 
\cite{Hannestad:2000gt}, in which the collision integral is taken to be $\hat C[\delta f] \simeq \delta f/\tau_c$, $\tau_c = \langle a n \sigma v \rangle^{-1}$ being the mean conformal time between collisions. 
We can rewrite the Boltzmann equation in a more convenient way (we refer the reader to Ref. \cite{Ma:1995ey} for the notation):
\begin{equation}
\frac{\partial \Psi_i}{\partial \tau} + i \frac{q(\vec{k} \cdot \hat{n})}{\epsilon} \Psi_i + \frac{d \ln f_0}{d \ln q} \left[ \dot{\eta} - \frac{\dot{h}+6\dot{\eta}}{2} \left(\hat{k} \cdot \hat{n} \right)^2 \right] = - \Gamma_{ij} \Psi_j\, ,
\end{equation}
where the indices $i$ and $j$ label neutrino mass eigenstates, and summation over repeated indices should be understood.
In the case under consideration, the scattering cross section $\sigma$ is of the order of $G_X^2 T_\nu^2$, where $T_\nu = (3/11)^{1/3} T_\gamma$ is the common temperature of active and sterile neutrinos after flavour equilibration. Given that the neutrino number densities $n_s = n_\nu = (3/2) (\zeta(3)/\pi^2)\, T_\nu^3$, we have that the collision rate $\Gamma = \tau_c^{-1} \sim a\,G_X^2 T_\nu^5$. Comparing this with the conformal Hubble expansion rate $\mathcal{H}\equiv aH$, we can find the time at which collisions cease to be important and sterile neutrino start to behave as free-streaming particles. 

\begin{figure}[h!]
\includegraphics[scale=0.25]{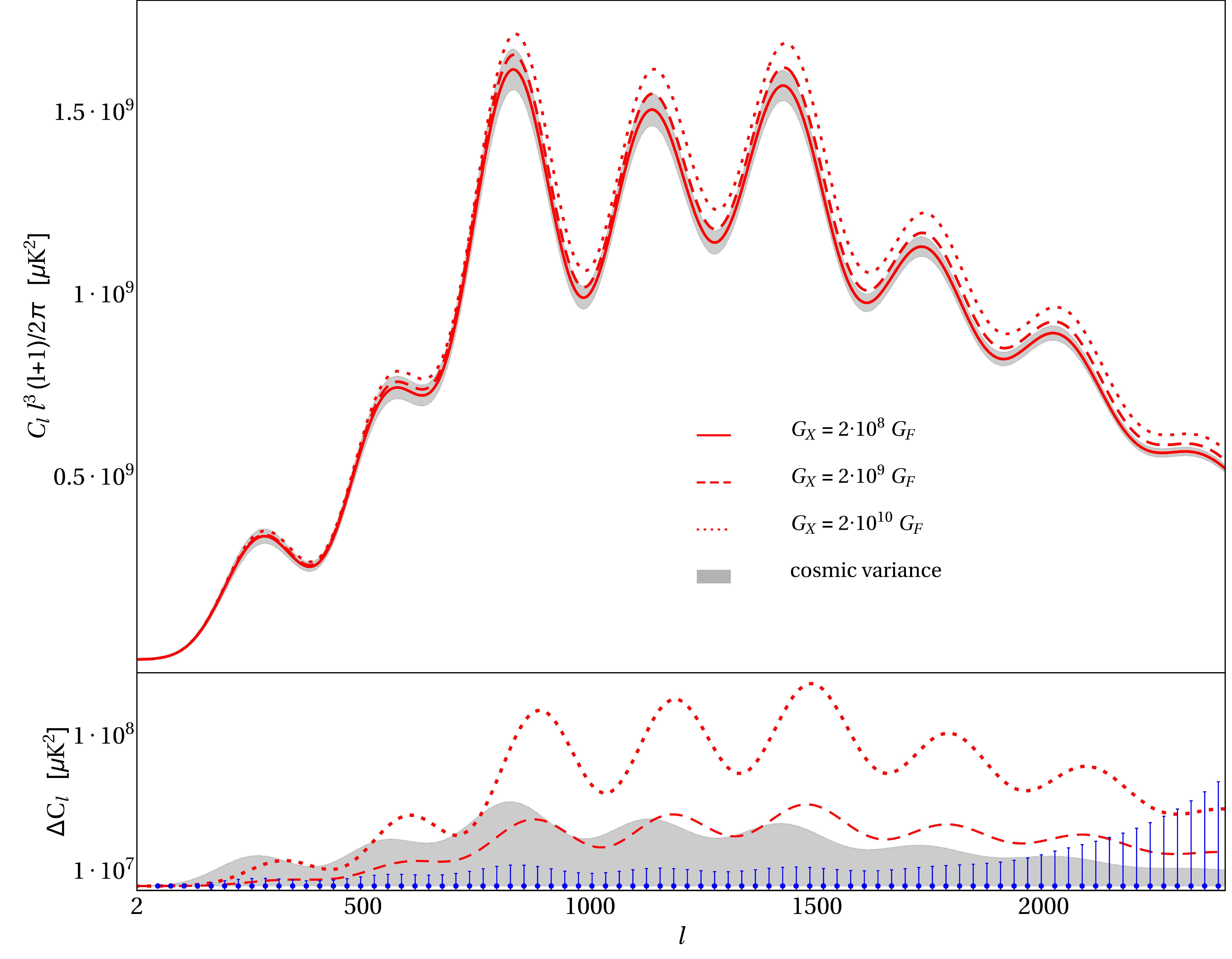}
\caption{Angular power spectrum of CMB temperature fluctuations. For the non interacting case and in case of secret interactions. In the upper panel, we show spectra for three different values of the coupling constant $G_X = 2\times \left\{10^{8},\, 10^9,\,10^{10}\,\right\}G_F$ (red solid, dashed, dotted lines, respectively). The non-interacting case is undistinguishable from the $G_X = 2\times 10^{8}\,G_F$ case. The APS is obtained assuming 3 active neutrinos having $\sum_{i=1}^3 m_{i} = 0.06\, \mathrm{eV}$ and a sterile neutrino species with $m_s = 1\, \mathrm{eV}$. In the lower panel, we show residuals with respect to the non-interacting case. The error bars represent the uncertainties of the Planck 2015 data.}
\label{fig:ps_int}
\end{figure}

Boltzmann codes like \texttt{camb} \cite{Lewis:1999bs} evolve the perturbations in the distribution functions of the mass eigenstates. In order to obtain the scattering rates between mass eigenstates, those should be projected from
the flavour basis through the mixing matrix.  We shall assume that the sterile state is the superposition of the 1 and 4 mass eigenstates through the vacuum mixing 
angle $\theta_s$ as 
\begin{equation}
\nu_s \simeq \sin \theta_s \nu_1 + \cos \theta_s \nu_4 \,\ ,
\label{eq:mixing}
\end{equation}
so that we are in the situation in which the mass eigenstates $\nu_1$ and $\nu_4$ interact with relative rates 
$\sin^2 \theta_s$ and $\cos^2 \theta_s$, while $\nu_2$ and $\nu_3$ are essentially free-streaming \cite{Chu:2015ipa}, and the scattering rate term becomes:
\begin{equation}
\Gamma_{ij} = 
\begin{bmatrix}
\sin^2\theta_s & 0 & 0 & \sin \theta_s\, \cos \theta_s\\ 
0 & 0 & 0 & 0 \\ 
0 & 0 & 0 & 0 \\ 
\sin \theta_s\, \cos \theta_s & 0 & 0 & \cos^2 \theta_s
\end{bmatrix}(3/2) (\zeta(3)/\pi^2)\, a G_X^2\, T_\nu^5\, .
\label{eq:matrix}
\end{equation}
It is possible to rewrite the Boltzmann equation for the mass eigenstates as an infinite hierarchy of multipoles \cite{Ma:1995ey}:
\begin{subequations}
\begin{align}
&\dot{\Psi}_{i,0} = -\frac{q k}{\epsilon} \Psi_{i,1} + \frac{1}{6} \dot{h} \frac{d \ln{f_0}}{d \ln{q}} \, ,\\
&\dot{\Psi}_{i,1} = \frac{qk}{3 \epsilon} \left(\Psi_{i,0} - 2 \Psi_{i,2} \right) \, ,\\ 
&\dot{\Psi}_{i,2} = \frac{qk}{5 \epsilon} \left( 2 \Psi_{i,1}- 3 \Psi_{i,3} \right) - \left( \frac{1}{15} \dot{h} + \frac{2}{5}\dot{\eta} \right) \frac{d \ln{f_0}}{d \ln{q}} - \Gamma_{ij} \Psi_{j,2}\, ,\\
&\dot{\Psi}_{i,\ell} = \frac{qk}{(2\ell+1)\epsilon} \Big[ \ell \Psi_{i,(\ell-1)} - (\ell+1) \Psi_{i,(\ell+1)} \Big]  - \Gamma_{ij} \Psi_{j,\ell} \quad (\ell\ge 3)\, ,
\end{align}
\label{eq:boltz_mass_int}
\end{subequations}
where $\ell$ is the parameter of the Legendre expansion.
We have set to zero the $\ell=0$ and $\ell=1$ terms of collision integral, in order to ensure particle number  and momentum conservation. Thus the scattering directly affects the neutrino fluid from the shear onwards, and then propagates to the lower order moments.
It is clear that, as long as the collision rate is much larger than the expansion rate, interacting neutrinos behave as perfect fluid.\footnote{In this case, the system of Eqs.
(\ref{eq:boltz_mass_int}) becomes stiff and a direct numerical integration is unfeasible. During this tight-coupling regime, we 
only evolve the $\ell = 0,\,1$ moments of the distribution, using an approximate form for $\ell=2$ to close the hierarchy (see e.g. Ref. \cite{Cyr-Racine:2013jua}).}This means 
that shear and higher moments are exponentially suppressed, and the power in fluctuations is bound to the local monopole and dipole of the neutrino fluid. The net effect is that, at scales that are within the horizon during the interacting regime, density and pressure perturbations are enhanced with respect to the non-interacting case. This enhancement propagates to the photon fluid, and thus to CMB anisotropies, through the metric perturbations, as it can be clearly seen in Fig. \ref{fig:ps_int}, 
where we plot the temperature angular power spectrum (APS) (multiplied by an additional factor of $\ell^2$) for three models with an interacting sterile neutrino with $m_s = 1\, \eV$ and different values of the coupling strength  $G_X$. In all cases we take $\neff = 2.7$, consistently with the expectation of flavour equilibration. The prediction for the case with  $G_X \sim 10^8 G_F\simeq 10^3\, \GeV^{-2}$
 is practically identical to that of a \LCDM\ extension with one non-interacting sterile neutrino and $\neff = 2.7$. This means that there is a range of values around $G_X\sim 10^8 G_F$ in which we still have a copious production of sterile neutrinos and flavour equilibration, but no direct effects of the interaction are visible on the APS (still, the effective number of relativistic degrees of freedom is $\neff=2.7$).
Larger values of
$G_X$ change the spectrum by increasing the power below a critical scale, related to the size of the horizon at the time at which neutrinos enter the free-streaming
regime. For the parameter values used in the plot, we have that the comoving scale that enters the horizon at this time is $k \simeq 0.01\,\Mpc^{-1}$ ($k \simeq  0.03 \,\Mpc^{-1}$), mapped to $\ell \simeq 130$ ($\ell \simeq 400$) for $G_X=10^9 G_F$ ($G_X=10^{10} G_F$).
In the following we will also consider BAO data to derive constraints on the parameters of the model. It is known that the inclusion of BAO measurements greatly improves constraints on neutrino masses, due to the breaking of geometrical degeneracies (most importantly, the one between the mass and $H_0$). As we shall see, BAO data also help in constraining the strength of secret interactions. Even if the effect of secret interactions is only seen at perturbation level, while BAO probe the background expansion, nevertheless they help to break parameter degeneracies that are present when only CMB are considered. In particular, the effect of a delayed neutrino decoupling can be canceled by increasing the total matter density, since the lesser amount of early-integrated Sachs-Wolfe effect will compensate the enhancement of perturbation power described above. We thus expect that constraining the matter density through BAO observations will result in tighter constraints on $G_X$.

\section{Cosmological analysis}
In this section we discuss our analysis of available cosmological data to constrain the coupling of the secret interaction. We first describe the method and data used, and then we present our results.

\subsection{Method and data}

We use the Boltzmann code \texttt{camb} \cite{Lewis:1999bs}, modified as described in the previous section, to follow the evolution of cosmological perturbations and compute the CMB anisotropy power spectra for given values of the cosmological parameters, including the secret coupling $G_X$ and the mass $m_s\equiv m_4$ of the (mostly) sterile neutrino. 
In order to derive Bayesian credible intervals for the parameters, we use the Markov Chain Monte Carlo (MCMC) engine
\texttt{CosmoMC}   \cite{Lewis:2002ah} (interfaced with the modified \texttt{camb}). Our parameter space consists of the six parameters of the $\Lambda$CDM model, augmented
by the parameters describing the sterile neutrino sector. The $\Lambda$CDM parameters are the baryon and cold dark matter densities $\omega_b\equiv \Omega_b h^2$ and $\omega_c \equiv \Omega_c h^2$, the angle $\theta$ subtended by the sound horizon at recombination, 
the optical depth to reionization $\tau$ and the logarithmic amplitude $\ln(10^{10}A_s)$ and spectral index $n_s$ of the primordial spectrum of scalar fluctuations.
The neutrino sector is instead described by the secret coupling $G_X$ and the sterile neutrino mass $m_s$. As explained in the previous section,
we fix $\neff = 2.7$, consistently with the assumption that all neutrino states (both active and sterile) have a common temperature $T_\nu =  (3/11)^{1/3} T_\gamma$. This amounts to the assumption that $G_X \gtrsim 10^8 G_F$. We also fix the active-sterile mixing angle to $\theta_s = 0.1$ and the sum of the masses of the (mostly) active neutrinos to $0.06\,\eV$, equally shared among three mass eigenstates. We further assume flat spatial
geometry and adiabatic initial conditions. 
  
In our analysis, we always take flat, wide (in the sense that they are much larger than the expected posterior widths) priors for the six \LCDM\ parameters. We also consider priors on $G_X$ and $m_s$ in order to model limiting cases of the scenario under consideration, to include additional pieces of experimental information, or simply to explore
different regions of the parameter space. We start by performing a set of exploratory MCMC runs in which we assume a flat prior distributions in $\log_{10}\left[G_X\right]$ and $m_s$. The advantage of a logarithmic prior in $G_X$ is that 
it allows to explore several orders of magnitude in the parameter with equal probability per decade and thus to assess when the
effect of secret interactions on the CMB APS becomes ``large'', at least in comparison with the experimental sensitivity. However,
a logarithmic prior gives more weight to small values of the parameter with respect to a flat prior, resulting in tighter bounds on the parameter itself. Moreover,
it is an improper prior, since it does not integrate to a finite value if $G_X\ge 0$, and in order to give meaningful credible intervals an arbitrary, non-zero, lower
bound on $G_X$ has to be assumed. For these reasons, we only use the results from this analysis to estimate the sensitivity of the data to $G_X$ and to gauge the initial step of the subsequent MCMC runs, that always use a flat prior on $G_X$.

The full model, in which the \LCDM\ parameters as well as $G_X$ and $m_s$ are varied, is dubbed \SLCDM\ (standing for ``\LCDM\ with secret interactions''). 
In this case, and unless otherwise stated, we take flat and wide priors also on $G_X$ and $m_s$.
Note that, as explained above, we always have $\neff = 2.7$.  
A limiting scenario is obtained by fixing $G_X$ to a very small value in our modified \texttt{camb} while keeping $\neff=2.7$,
in order to reproduce the case in which $G_X$ is large enough for the flavour equilibration to happen, while still being small enough not to affect the evolution of cosmological perturbations. As we have mentioned, this approximately corresponds to $G_X \sim 10^8 G_F$. Since, as noted in the previous section, this case is practically indistinguishable, as long as the evolution of cosmological perturbations in concerned, from a $\Lambda$CDM scenario with $\neff = 2.7$ and $G_X=0$, we shall refer to this model as ``\SLCDM\_GX0''.
Finally, we also consider prior on $m_s$ to model information from short baseline experiments. We refer to Ref. \cite{Gariazzo:2015rra} in which the allowed $3\sigma$ (i.e., 99.73\% 
CL) range for the squared mass difference $\Delta m^2_{41}=m_4^2- m_1^2$ that explains the SBL anomalies is $0.87\,\  \eV^2 \le \Delta m^2_{41} \le 2.04\,\ \eV^2$. Not knowing the full shape of the probability density distribution for $m_s$, 
we decided to model it considering two ``extreme'' cases: in the first (``narrow $m_s$ prior'') we impose a gaussian prior
$m_s = 1.27 \pm 0.03 \,\ \eV$ (the width of the prior is chosen to match the 1$\sigma$ confidence interval for $\Delta m^2_{41}$ \cite{Gariazzo:2015rra}, assuming $m_4 \gg m_1$), while in the second (``broad $m_s$ prior'') we impose a flat prior $0.93\,\ \eV \le m_s \le  1.43 \,\ \eV$,
corresponding to the $3\sigma$ interval reported above. 
Finally, we will often compare our results to those obtained in the framework
of the standard \LCDM\ model; for these, we refer to the values reported in the Planck 2015 parameters paper \cite{Ade:2015xua}, and in this case it should be understood that $\neff = 3.046$ \cite{Mangano:2005cc, deSalas:2016ztq}. A list of the abbreviations used for the models considered in this paper, including a short description, can be found in Tab. \ref{TAB:models}.

Our data consists of the baseline Planck 2015 dataset (dubbed ``PlanckTT+lowP'' in the Planck papers), that includes temperature data in the range $2\le \ell \le 2500$, as well as the large-scale
($2\le\ell\le 30$) polarization (based on the measurements of the 70 GHz channel) \cite{Aghanim:2015xee}. The likelihood function associated to the data
is computed using the code publicly released by the Planck collaboration\footnote{We acknowledge 
the use of the products available at the Planck Legacy Archive (\url{http://www.cosmos.esa.int/web/planck/pla}).}. We marginalize over a number
of nuisance parameters related to astrophysical foregrounds and instrumental uncertainties, as described in Ref. \cite{Aghanim:2015xee}.
We also consider geometrical information coming from baryon acoustic oscillations; in particular we make use of the BAO results
from the 6dF Galaxy Survey  \cite{Beutler:2011hx}, from the BOSS DR11 LOWZ and CMASS samples \cite{Anderson:2013zyy}, and from the Main Galaxy Sample of the Sloan Digital Sky Survey   \cite{Ross:2014qpa}.
The extended dataset combining the Planck 2015 data with the BAO information will be denoted ``PlanckTT+lowP+BAO".

\begin{table}[h!]
\begin{center}
\begin{tabular}{CG}
 \hline\hline
 & Description \\
 \hline
 \LCDM & Standard six-parameter \LCDM, $\neff=3.046$. \\
 \SLCDM\_$\mathrm{GX0}$ & Sterile neutrino extension, $\neff=2.7$, $m_s$ free, ``small'' $G_X$  ($\sim 10^8 G_F$).  \\
 \SLCDM & Sterile neutrino extension, $\neff=2.7$, $m_s$ and $G_X$ free. \\
 \SLCDM\_Narrow & Sterile neutrino extension, $\neff=2.7$, $G_X$ free, $m_s = 1.27\pm0.03\, \eV$ (gaussian prior).\\
 \SLCDM\_Broad & Sterile neutrino extension, $\neff=2.7$, $G_X$ free, $0.93\,\ \eV \le m_s \le  1.43 \,\ \eV$ (flat prior).\\
 \hline\hline
\end{tabular}
\caption{Description of the models considered in this work.}
\label{TAB:models}
\end{center}
\end{table}

\subsection{Results}

We are now ready to present our results, summarized in Tabs. \ref{TAB:1} and \ref{tab:BAO}, where we show the Bayesian credible intervals for the parameters, for the various models and dataset combinations under consideration. As seen above, the presence of an interacting sterile neutrino impacts the cosmological observables in three ways:
\begin{itemize}
\item smaller $\neff$ due to flavour equilibration;
\item larger density of (possibly) free-streaming species;
\item reduced shear in the neutrino component of the cosmological fluid.
\end{itemize}
We start by considering the limit of small $G_X$ ($\sim 10^8 G_F$),  in which the third effect listed above is negligible, in order
to disentangle the first two effects. Comparing the columns for \LCDM\  and \SLCDM\_${\mathrm{GX0}}$, we note that there are considerable shifts in some parameters, in particular $H_0$ and $n_s$. 
The direction of the shifts is consistent with what we would expect given the well-known degeneracies of these parameters with both $\neff$ and the total density in light species. Looking at the $\chi^2$ values for the best-fit models, reported in Tab. \ref{TAB:2}, we find that \SLCDM\_${\mathrm{GX0}}$ performs worse in terms of goodness-of-fit, with a $\Delta\chi^2 = 7.7$ with respect to \LCDM. This is due to the low value of $\neff$ imposed by the flavour equilibration, while
Planck data prefer a value closer to the standard expectation $\neff=3.046$. We also note that the mass of the sterile is constrained to be $m_s < 0.82\,\ \eV$ at 95\% CL, 
thus being in strong tension with the values suggested by the SBL anomalies.

\begin{figure}[h!]
 \centering
 \includegraphics[scale=0.38]{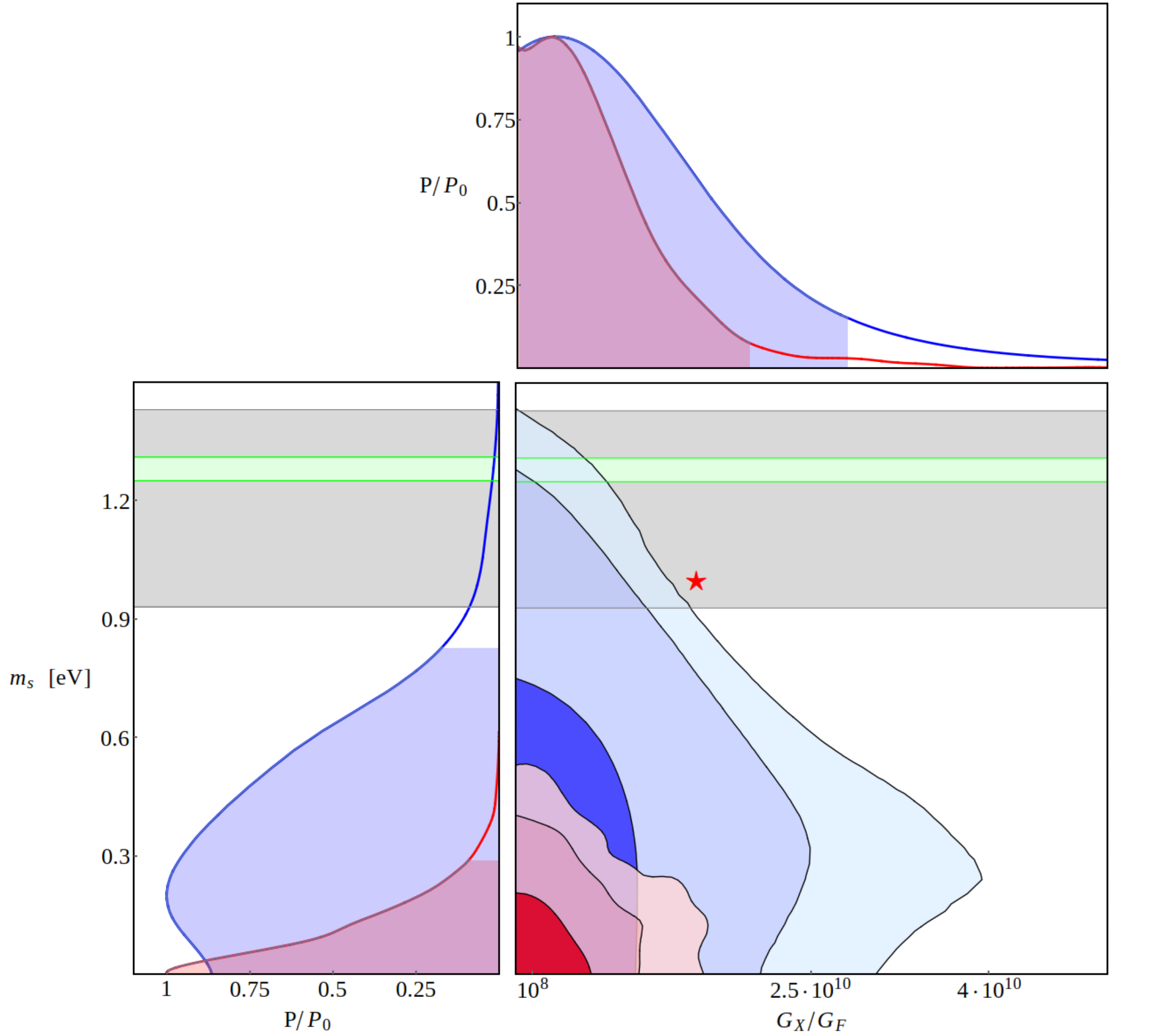}
 \caption{Two-dimensional (bottom right) and corresponding one-dimensional posteriors for the effective strength of the interaction $G_X = \sqrt{2} g_X^2 / 8 M_X^2$ in units of the Fermi constant (top) and the sterile neutrino mass $m_s$ (bottom left).
Blue constraints are obtained using PlanckTT+lowP data, while the red ones come from PlanckTT+lowP+BAO, both for the \SLCDM\ scenario (that assumes $G_X \ge 10^8 G_F$ and thus $\neff = 2.7$). The filled regions in the contour plot, from darker to lighter, show the 68, 95 and 99\% credible intervals. The shaded regions in the one-dimensional plots correspond to the 95\% credible interval. The grey and green horizontal regions are representative of the 68\% and 99.73\% priors on $m_s$ suggested by SBL anomalies. 
The red star at $G_X = 1.5\times10^{10} G_F$ and $m_s = 1\,\eV$ is representative of the strongly self-interacting scenario described in Refs. \cite{Dasgupta:2013zpn,Chu:2015ipa}. Note that the actual significance of the exclusion of the scenario with respect to \LCDM\ from the PlanckTT+lowP data is larger than $3\sigma$ (and similarly for the PlanckTT+lowP+BAO data), due to the fact that \LCDM\ does not belong to the parameter space shown in this figure (see discussion in the text).}
 \label{FIG:triangle}
\end{figure}

The impact of secret interactions can be assessed by varying the coupling strength as a free parameter of the model. To this purpose we compare \SLCDM\_${\mathrm{GX0}}$ to \SLCDM, shown in columns 2 and 3 of Tab.~\ref{TAB:1}. There are several points worth noticing: first of all, the constraints on the mass of the fourth eigenstate do not change with respect to the case of small $G_X$, thus remaining
in tension with the preferred SBL solution. Secondly, secret interactions stronger than $G_X = 2.8\times10^{10} G_F$ are disfavoured, precluding the possibility of the collisional regime lasting after $ z \sim \mathrm{few}\times 10^3$. Thus the scenario in the sterile neutrino starts to free stream
only after recombination, is disfavoured. This is consistent with the fact that the bound on $m_s$ that we get is of the same order of magnitude as the ones obtained by the Planck collaboration in a minimal extension of the \LCDM\ model. In that analysis, the effective mass $m_s^\mathrm{eff}\equiv 94.1\Omega_\nu h^2$ is used to parametrize the contribution of the sterile neutrino to the cosmological energy density. It is straightforward to see that, in our model, $m_s^\mathrm{eff}=(3/4)\, m_s$, so that in terms of the effective parameter the 95\% upper bound for \SLCDM\ reads $m_s^\mathrm{eff}<0.61\,\eV$. This should be compared with the result from the Planck collaboration for the same dataset (taken from the parameter tables available at the Planck Legacy Archive, see footnote 2), $m_s^\mathrm{eff}<0.88\,\eV$ at 95\% CL. 
The two values cannot be directly compared, since the Planck analysis considers $\neff$ as a free parameter, with a prior $\neff \ge 3.046$, while in our analysis it is fixed to $\neff = 2.7$. However, the tighter limit we find for $m_s^\mathrm{eff}$ is consistent with the lower value of $\neff$, given the direct correlation between the two parameters. In any case, the best-fit $\chi^2$ for \SLCDM\ is still worse than \LCDM\ ($\Delta\chi^2=3.9$) but yet better than \SLCDM\_${\mathrm{GX0}}$. 
When we also include information from BAO, we get tighter limits on $G_X$ and, especially, $m_s$, with 95\% upper bounds of $1.97 \times 10^{10}\,G_F$ and $0.29\,\eV$, respectively (see Tab.~\ref{tab:BAO}).
In Fig.~\ref{FIG:triangle}, we show the joint constraints and the marginalized one-dimensional posterior distributions for $G_X$ and $m_s$. For comparison,
in the two-dimensional plot, we also indicate with a red star a model with $G_X = 1.5\times 10^{10} G_F$ and $m_s = 1\,\eV$,
representative of the strong self-interacting scenario of Refs.~\cite{Dasgupta:2013zpn,Chu:2015ipa}, that was argued to reconcile cosmological measurements and sterile neutrino interpretation of SBL anomalies (note that the other scenario considered in Ref.~\cite{Chu:2015ipa}, with weak self-interactions, is not mapped by our analysis).
In particular, a value $G_X \sim 10^{10} G_F$ roughly corresponds to the white band in the upper left part of Fig. 4 of Ref.~\cite{Chu:2015ipa} (at least down to the point where the 4-point approximation is valid, namely $M_X\sim 10^{-2}$ MeV and $g_X\sim 10^{-3}$) and the red star in that figure corresponds to a model with $G_X = 1.5\times 10^{10} G_F$.
We stress that, even if from this figure the scenario considered in Refs.~\cite{Dasgupta:2013zpn,Chu:2015ipa} seems to be excluded at the $\sim 3\sigma$ level for our most conservative choice of the dataset, i.e. PlanckTT+lowP, and even more strongly for PlanckTT+lowP+BAO,
the actual statistical significance of the exclusion is somehow larger in both cases.  A proper assessment should take into account that models with sterile secret interactions with $G_X>10^8\,G_F$ have $\neff = 2.7$,
a value that is itself mildly disfavoured with respect to the \LCDM\ prediction of $\neff=3.046$. In the following paragraph, we will better quantify this statement, for the PlanckTT+lowP dataset, by comparing $\chi^2$ values between the best-fit models for \LCDM\ and \SLCDM.

In order to better assess the (dis)agreement between Planck CMB observations and the SBL anomalies, also in the presence 
of secret interactions, we look at the
fourth and fifth columns of Tab.~\ref{TAB:1}, where we show the results for the cases in which we impose priors on $m_s$ that mimic
the preferred SBL solution. 
For the \SLCDM\_Broad model (column 4 of Tab.~\ref{TAB:1}) we obtain almost the same constraint on the strength of the secret interaction we have obtained for the \SLCDM\ scenario, in spite of the larger value of $m_s$ imposed by the prior. We notice however that the posterior distribution for $m_s$ 
is peaked in the lower edge of the prior, $m_s = 0.93\, \eV$. In the \SLCDM\_Narrow analysis, on the other hand, the larger \textit{a priori} value of the sterile neutrino mass, $m_s\simeq 1.27$ eV, yields a looser constraint $G_X < 4\times 10^{10}\, G_F$.
For these two models, we see that the best-fit $\chi^2$ (computed on the PlanckTT+lowP dataset) is much worse than \LCDM: $\Delta\chi^2 = 11.1$
and $12.5$ for the ``broad'' and ``narrow'' priors, respectively. We argue that the inclusion of BAO information would make the tension even stronger,
given the preference of that dataset for small values of the sterile neutrino mass. Finally, we notice how all models with non-standard interactions show a preference for
values of $H_0$ even smaller than the one obtained in the framework of \LCDM\ (see corresponding row of Tab.~\ref{TAB:1}) further increasing
the tension between CMB and direct estimates of the Hubble constant \cite{Riess:2016jrr}; this is not captured 
by the $\chi^2$ figures reported above, that are computed on CMB data only. The increased tension is due in part to the low value of $\neff$,
and, in the models with the SBL priors, by the large value of $m_s$; both effects, as per known degeneracies, push towards a smaller $H_0$.

\begin{table}[h!]
\resizebox{\textwidth}{!}{
\begin{tabular}{MLLLLL}
\hline\hline
Parameter & \LCDM\ & \SLCDM\_GX0 & \SLCDM & \SLCDM\_Broad & \SLCDM\_Narrow  \\
\hline
$\Omega_b h^2$ & $0.02222\pm0.00023$ & $0.02177\pm0.00024$ & $0.02172\pm0.00025$ & $0.02167\pm0.00025$ & $0.02166^{+0.00024}_{-0.00024}$ \\
$\Omega_c h^2$ & $0.1197\pm0.0021$ & $0.1167\pm0.0022$ & $0.1171\pm0.0023$ & $0.1165\pm0.0022$ & $0.1160\pm0.0021$ \\
$100 \theta_{MC}$ & $1.04085\pm0.00047$ & $1.04103\pm0.00050$ & $1.04323^{+0.00091}_{-0.00073}$ & $1.04319\pm0.00074$ & $1.04307^{+0.0010}_{-0.00077}$ \\ 
$\tau$ & $0.078\pm0.019$ & $0.070\pm0.018$ & $0.065\pm0.018$ & $0.067\pm0.018$ & $0.066\pm0.018$ \\ 
$n_s$ & $0.9655\pm0.0061$ & $0.9448\pm0.0070$ & $0.9284\pm0.0088$ & $0.9191^{+0.0076}_{-0.0078}$ & $0.9161^{+0.0081}_{-0.0072}$ \\ 
$\ln(10^{10}A_s)$ & $3.089\pm0.036$ & $3.063\pm0.035$ & $3.023\pm0.038$ & $3.027\pm0.037$ & $3.028\pm0.036$ \\ 
\hline
$G_X/G_F$ & -- & $10^8$ & $ <2.8 \times10^{10}$ & $ < 2.9\times10^{10}$ & $< 4.0\times10^{10}$ \\  
$m_s$& -- & $< 0.82$& $ < 0.82$ & $[0.93 , 1.30]$ &$1.27 \pm 0.028$ \\
\hline
$H_0$ & $67.31\pm0.95$ & $62.2^{+2.0}_{-1.7}$ & $62.6^{+1.8}_{-1.8}$ & $59.56\pm0.88$ & $58.91^{+0.82}_{-0.79}$ \\  
\hline \hline
\end{tabular}}
\caption{Parameter constraints for the models under consideration, from the PlanckTT+lowP dataset. We either quote constraints in the form ``mean $\pm$ 68\% uncertainty'', or as 95\% credible intervals (when not indicated, the lower limit should be understood to be zero). Units of $m_s$ and $H_0$ are eV and km s$^{-1}$ Mpc$^{-1}$, respectively.}
\label{TAB:1}
\end{table}

\begin{table}[h!]
\begin{center} 
\begin{tabular}{ML}
\hline\hline
Parameter & S$\Lambda$CDM  \\ 
\hline
$\Omega_b h^2$ & $0.02197\pm0.00021$ \\
 $\Omega_c h^2$ & $0.1144^{+0.0016}_{-0.0015}$ \\ 
 $100 \theta_{MC}$ & $1.04332^{+0.00090}_{-0.00063}$ \\ 
 $\tau$ & $0.074\pm0.018$ \\ 
 $n_s$ & $0.9392\pm0.0063$ \\ 
 $\ln(10^{10}A_s)$ & $3.038\pm0.036$ \\
 \hline
 $G_X/G_F$ & $<1.97\times10^{10}$ \\ 
 $m_s$ & $<0.29$\\
 \hline 
 $H_0$ & $65.26\pm0.68$ \\ 
 \hline\hline
\end{tabular}
\caption{Parameter constraints for the models under consideration, from the PlanckTT+lowP+BAO dataset. We either quote constraints in the form ``mean $\pm$ 68\% uncertainty'', or as 95\% credible intervals (when not indicated, the lower limit should be understood to be zero). Units of $m_s$ and $H_0$ are eV and km s$^{-1}$ Mpc$^{-1}$, respectively. \label{tab:BAO}}
\end{center}
\end{table}

\begin{table}[h!]
\small
\begin{center} 
\resizebox{\textwidth}{!}{%
\begin{tabular}{SMMMMM}
\hline\hline
Parameter & \LCDM &\SLCDM\_GX0 & \SLCDM & \SLCDM\_Broad & \SLCDM\_Narrow\\ 
\hline
$\chi^2_\mathrm{min}$ & $11265.1$ & $11272.8$ & $11269.0$& $11275.2$& $11277.6$ \\
\hline\hline
\end{tabular}}
\caption{Best-fit $\chi^2$ values for the models under consideration, for the PlanckTT+lowP dataset.}
\label{TAB:2}
\end{center}
\end{table}

\begin{figure}[h!]
 \centering
 \includegraphics[scale=0.38]{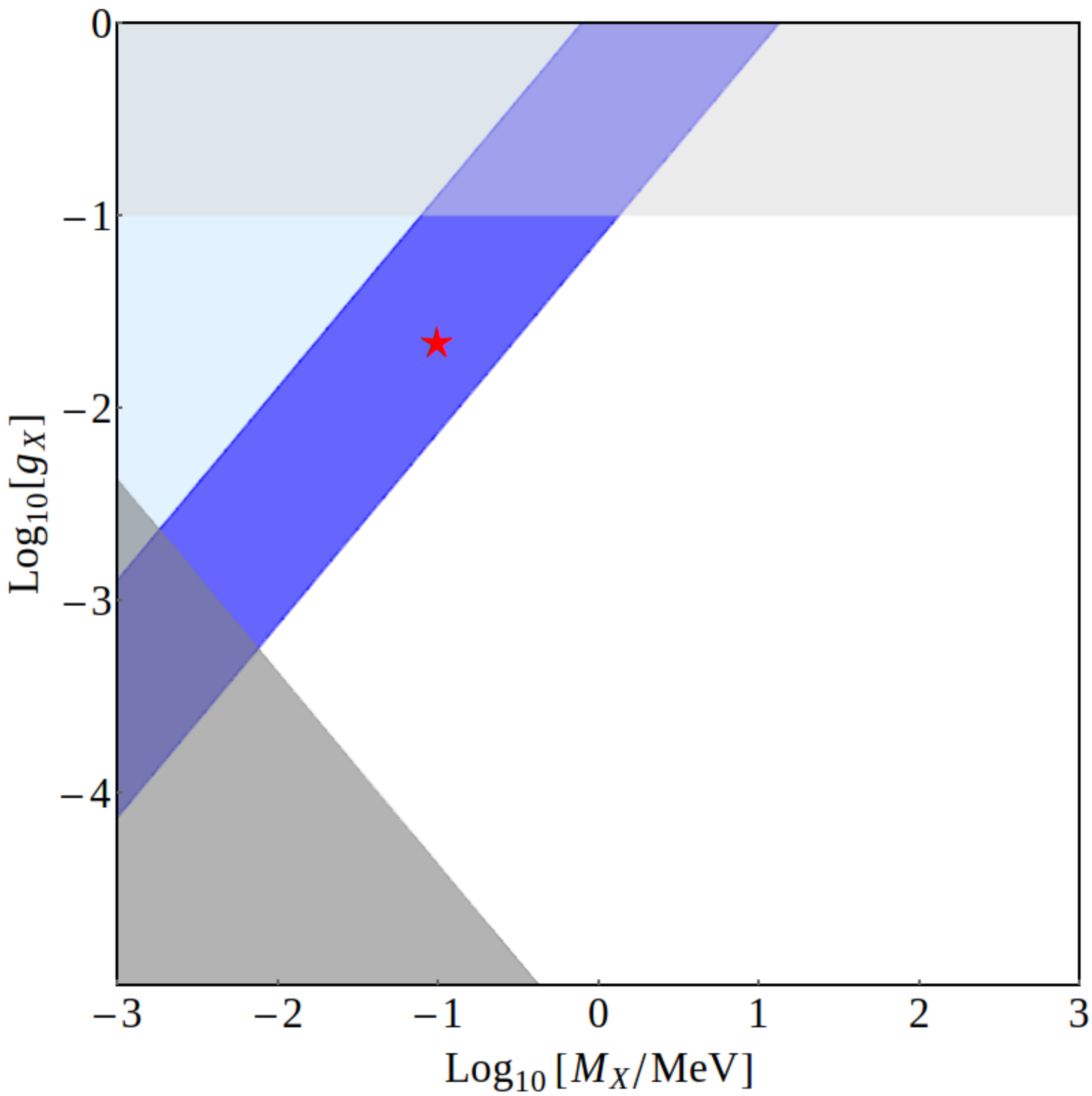}
 \caption{Two-dimensional allowed parameter space for the dimensionless coupling constant $g_X$  and the mediator mass $M_X$. The light and dark blue areas show the region excluded by this study. The light blue region corresponds to values of the interaction strength $G_X > 2.9 \times 10^{10}\, \GeV^{-2}$, thus larger than the $95\%$ upper limit on this parameter from \planck. In the dark blue region $10^8\, \GeV^{-2} < G_X < 2.9 \times 10^{10}\, \GeV^{-2}$, but is still disfavoured as it does not allow to circumvent the neutrino mass bound. The regions where the approximations used in our study become to break down are  colored in gray: the light gray band on top indicates the non-perturbative regime ($g_X \gtrsim 0.1$) while the dark gray triangle on the bottom-left is where the interaction cannot be described as four-point. The red star is representative of the strongly self-interacting scenario described in Refs. \cite{Dasgupta:2013zpn, Chu:2015ipa}.}
\label{2dplot}
\end{figure}

\section{Conclusions}

In this work we have investigated, using Planck 2015 observations and a compilation of BAO data, the feasibility of cosmological models with sterile neutrinos,
in addition to the three active states of the standard model of particle physics, with new, secret self-interactions mediated by a massive vector boson and confined in the sterile sector.
This model has been proposed in order to alleviate the tension between the preferred solution of the SBL neutrino anomalies and cosmological observations, that disfavour a fourth fully thermalized neutrino species.
Notably  the effect of the new interactions would be to effectively dilute the density of both the active and sterile states (leading to an effective number of relativistic species $\neff=2.7$, more compatible with the Planck data). However, the mass of the sterile neutrino required to explain the SBL anomalies still appears to be too large with respect to the corresponding cosmological bounds. It was not clear a priori if and to what extent such bounds could be evaded 
thanks to the secret interactions that, if very strong, could significantly delay the onset of sterile neutrino free streaming.

Secret interactions also leave an imprint on the CMB spectra, by extending the collisional regime for the neutrino fluid. Using this effect, we have constrained the effective ``Fermi constant'' $G_X$ of the new interaction to be smaller than $2.8 \times 10^{10}\, G_F$ at 95\% CL from the Planck 2015 temperature and large-scale polarization data. This limit is improved to $2.0 \times 10^{10}\, G_F$ at 95\% CL when information from BAO are included.
These results disfavour the range, corresponding to $G_X \gtrsim 10^{10} G_F$, in which the onset of sterile neutrino free streaming is delayed until after recombination, and cosmological mass bounds could be possibly evaded.
In fact, our self-consistent analysis yields, at 95\% CL, $m_s < 0.82\,\eV$ and $m_s < 0.29\,\eV$ from the Planck 2015 data alone and in combination with BAO, respectively, smaller than the value required to explain SBL anomalies, allowing
to conclude that the tension between the SBL oscillation experiments and CMB observations still holds even in extended models with secret sterile neutrino interactions.
Even disregarding BAO data, secret interactions with $G_X \gtrsim 10^8 G_F$ 
are disfavoured with respect to standard \LCDM, by CMB data, due to their prediction of a low $\neff$. Moreover,
CMB estimates of the Hubble constant $H_0$ in the secret interactions framework are smaller than their \LCDM\ counterparts,
thus increasing the tension with astrophysical measurements of the same quantity.

We summarize our findings in Fig. \ref{2dplot}, where we show the parameter space excluded by our analysis in terms of the dimensionless coupling constant $g_X$ and the mediator mass $M_X$. The excluded region coincides with the whole region in which our assumptions hold and the approximations used are valid. 
The light and dark blue areas show the region excluded by our work. In particular,  the light blue region corresponds to values of the interaction strength $G_X > 2.9 \times 10^{10}\, \GeV^{-2}$, thus larger than the $95\%$ upper limit on this parameter from \planck. In the dark blue region the range  $10^8\, \GeV^{-2} < G_X < 2.9 \times 10^{10}\, \GeV^{-2}$ is still disfavoured by  the neutrino mass bound.
The red star is representative of the strong self-interacting scenario described in Refs. \cite{Dasgupta:2013zpn,Chu:2015ipa}.
 The regions where the approximations used in our study become to break down are  colored in gray. The horizontal band in  light gray band on top indicates the non-perturbative regime ($g_X \gtrsim 0.1$) while the dark gray triangle on the bottom-left is where the interaction cannot be described as four-point interaction. This is obtained when the temperature at which $\nu_s$ are produced
 (approximated with Eq.~(12) of \cite{Saviano:2014esa})
  is comparable or larger than the mediator mass $M_X$. 

Our analysis has excluded the possibility of a single sterile neutrino with $\sim 1$ eV mass and $\sim 0.1$ mixing (as preferred by the SBL anomalies) with active neutrinos, having strong, four-fermion pointlike self-interactions. This is because it is not possible to hide the cosmological effects of such a large neutrino mass by means of a reduced free-streaming, without at the same time injecting too much extra power in the CMB angular power spectra.
As it can be seen by comparing our Fig. 3 with Fig. 4 of Ref. \cite{Chu:2015ipa} (please note that the quantities reported on the vertical axes of the two figures are related by $\alpha_s = g_X^2/4\pi$), the present analysis excludes the thin white band in the upper left part of Fig. 4 of Ref.  \cite{Chu:2015ipa} (dubbed there ``strong self-interactions'' region), that was regarded as being of particular interest as it could help explain the problems that arise at small scales in cold dark matter models of structure formation. On the other hand, the region in the lower part of Fig. 4 of Ref  \cite{Chu:2015ipa}, corresponding to weak self-interactions, is not probed by our analysis. Even if a first exploratory study indicated this region  as possible solution of the light sterile neutrino problem, recent refined calculations show that also this possibility seems to be ruled out (see Refs \cite{Cherry:2016jol}, and \cite{Chu:inprep}), due to the X-mediated s-channel process leading to efficient sterile neutrino production. To conclude, we remark that our analysis assumes that the mass of the mediator is much larger than the temperatures relevant for the problem, and that self-interactions are perturbative. Moreover, we have not considered the case of two or more species of sterile neutrinos \cite{Tang:2014yla}, although we argue that, in the case of complete thermalisation, they would be even more in tension with observations due to i) an even lower value of $\neff$, and ii) a larger density of interacting species, presumably resulting in a stronger bound on $G_X$.
  

\section*{Acknowledgments}

We thank Basudeb Dasgupta and Joachim Kopp for reading our manuscript and for valuable comments on it.
This paper is based on observations obtained with the satellite \planck\ (\url{http://www.esa.int/Planck}),
an ESA science mission with instruments and contributions directly funded by ESA Member
States, NASA, and Canada.
We acknowledge support from ASI through ASI/INAF Agreement 2014-
024-R.1 for the Planck LFI Activity of Phase E2.
We acknowledge the use of computing facilities at NERSC (USA). We acknowledge 
the use of the products available at the Planck Legacy Archive. The work of G.M. is supported by the Istituto Nazionale di Fisica Nucleare (INFN) through the ``Theoretical Astroparticle Physics'' project.
The work of A.M. is supported by the Italian Ministero dell'Istruzione, Universit\`a e Ricerca (MIUR) and Istituto Nazionale di Fisica Nucleare (INFN) through the ``Theoretical Astroparticle Physics'' project. The work of N.S. has been supported by the German Research Foundation (DFG) under Grant Nos. KO 4820/1Ð1, FOR 2239, EXC-1098 (PRISMA) and by the European Research Council (ERC) under the European UnionÕs Horizon 2020 research and innovation programme (grant agreement No. 637506, Ò$\nu$DirectionsÓ).

\end{document}